\documentclass[preprint]{iucr}
\journalcode{S}

\usepackage{amssymb}
\usepackage{amsmath}
\usepackage{array}
\usepackage{xcolor}
\usepackage{graphicx}
\usepackage{dcolumn}
\usepackage{bm}
\graphicspath{ {Figures/} }
\usepackage[ruled,vlined]{algorithm2e}

\newcommand{\change}[1]{{\color{black}#1}}

\begin{document}

\title{Quantitative analysis of diffraction by liquids using a pink-spectrum X-ray source}

\cauthor[a]{Saransh}{Singh}{saransh1@llnl.gov}{address if different from \aff}
\author[a]{Amy L.}{Coleman}
\author[b]{Shuai}{Zhang}
\author[a]{Federica}{Coppari}
\author[a]{Martin G.}{Gorman}
\author[a]{Raymond F.}{Smith}
\author[a]{Jon H.}{Eggert}
\author[a]{Richard}{Briggs}
\author[a]{Dayne E.}{Fratanduono}

\aff[a]{Lawrence Livermore National Lab, Computational Engineering Division, Livermore, CA 94511 \country{USA}}
\aff[b]{Laboratory for Laser Energetics, University of Rochester, Rochester, NY 14623, USA}

\shortauthor{Singh et al.}

\maketitle 


\begin{synopsis}
This paper describes a new approach to compute structure factors, radial distribution function and mean density from liquid scattering data in a pink x-ray source.
\end{synopsis}

\begin{abstract}
We describes a new approach for performing quantitative structure-factor analysis and density measurements of liquids using x-ray diffraction with a pink-spectrum x-ray source. The methodology corrects for the pink beam effect by performing a Taylor series expansion of the diffraction signal. The mean density, background scale factor, peak x-ray energy about which the expansion is performed, and the cutoff radius for density measurement are estimated using the derivative-free optimization scheme. The formalism is demonstrated for a simulated radial distribution function for tin. Finally, the proposed methodology is applied to experimental data on shock compressed tin recorded at the Dynamic Compression Sector at the Advanced Photon Source, with derived densities comparing favorably to other experimental results and the equations of state of tin.
\end{abstract}

\section{Introduction}
Over the last two decades marked improvements have been made to both the experimental and analytical techniques associated with the study of dense liquid states using x-ray diffraction and static high-pressure techniques \cite{Eggert2002,Morard2014}. Typically, the study of high-pressure liquids (P $<$ 100 GPa) have relied on the diamond anvil cell (DAC) apparatus, which consists of two opposing diamond anvils that compress a sample (surrounded by a pressure-transmitting medium) in a metallic chamber compressed between the two anvils. While much important research has been conducted using static compression techniques at relatively low pressures, e.g. the observation of first order liquid-liquid phase transitions \cite{Katayama2000,Soper2008},  and critical point in sulphur near 2.0 GPa \cite{Henry2020}, there is an inherent limit to the accessible pressure states, imposed by the strength of the diamonds in the cell, as well as a upper temperature limit imposed by the physical geometry of DAC apparatus \cite{Anzellini2020}. Additionally, the thick diamonds used in the DAC setup are known to make a significant contribution to the x-ray diffraction signal collected during high-temperature, high-pressure liquid experiments, along with scattering contributions from the surrounding medium. The removal of these parasitic features can be non-trivial and is essential for the proper analysis of the x-ray scattering from the material of interest \cite{Eggert2002,Morard2014}. The recent implementation of Soller slits has facilitated the collection of high quality diffraction data from low Z liquids at pressures just over 1 Mbar \cite{Weck2017}, however small sample sizes and the practical difficulties of using DACs at these conditions mean that it has so far been impossible to access the multi-megabar regime for these types of experiments (P $\textgreater$ 200 GPa).

\begin{figure}
    \centering
    \includegraphics[width=\columnwidth]{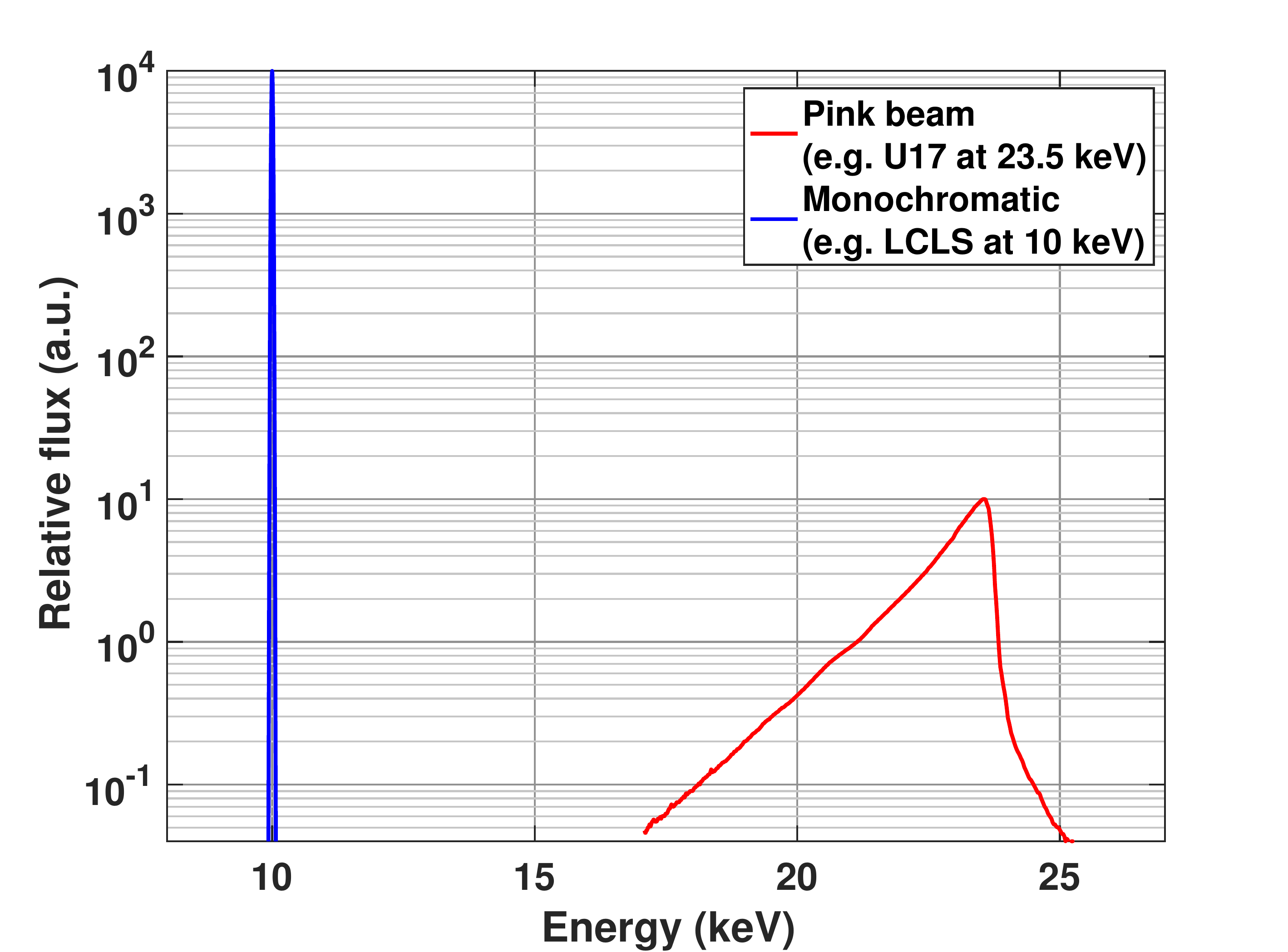}
    \caption{Representative (not measured) x-ray photon energy-flux at the Linac Coherent Light Source (blue) and U17 flux measured at the Dynamic Compression Sector (red).}
    \label{fig:xrays}
\end{figure}

The advent of fourth generation light sources such as the Linac Coherent Light Source (LCLS) presents a new method of probing dense liquid states as generated through laser-driven dynamic compression experiments. The short timescales of such experiments require highly brilliant x-rays in order to obtain single-exposure diffraction data of high enough quality to perform quantitative analysis of liquid scattering data \cite{Briggs2019}. These shock-compression experiments grant access to pressure states up to several Mbar, vastly broadening the scope of the study of dense liquids. Liquid diffraction data has successfully been collected at LCLS \cite{Briggs2017,Gorman2018,Coleman2018}, however the detector coverage and the accessible momentum transfer or $q$-range over which diffraction data may be obtained can be limited. 
Recent years have seen the addition of such laser systems to synchrotron beamlines, meaning that laser-driven dynamic compression experiments can now also be conducted at facilities that have produced high quality liquid diffraction data from DACs over the previous two decades. These facilities have the capability to collect diffraction on sub-nanosecond timescales as well as providing detector coverage from a single panel detector for full azimuthal coverage making them excellent candidates for probing dynamically compressed liquids. The dynamic compression sector (DCS) at the Advanced Photon Source (APS), having recently installed a 100 J laser system, is one such facility that affords users the opportunity to dynamically compress samples \cite{DCS}. This beamline is equipped with a U17 undulator, providing a non-monochromatic x-ray source, commonly referred to as a pink beam. A representative x-ray photon flux vs energy curve for the x-ray free electron source at LCLS and the as measured spectral flux from the U17 undulator at APS is shown in Fig.~\ref{fig:xrays}. The full width at half maxima (FWHM) for the energy-flux distribution at DCS is $\sim 0.785$ keV, which is a $3.3\% $ spread around the energy of peak flux. The FWHM at LCLS is $\sim 0.2\%$. It is also worth noting that the relative flux at LCLS-II is about three orders of magnitude brighter compared to DCS.

The scattering of x-rays from a liquid sample produces a broad diffuse signal that contains information on short range order of the atoms. The shape and location of the broad liquid peaks are affected by the x-ray source type. A monochromatic x-ray beam (with bandwidth $\Delta$E/E $\ll 1$ \%) will typically produce peaks at a scattering angle defined by the average atomic positions. However, if the x-ray source is a pink-beam source, with an intensity profile characterized by a \change{sharp} Gaussian fall-off at higher energies and an exponential tail to lower energies with $\Delta$E/E $\sim$ 3\% (as is the case for the U17 undulator at DCS), then there is an artificial shift of the liquid peak locations to higher scattering angles and an asymmetry in the peak profile \cite{Bratos2014}. Analysis of the liquid scattering intensities as a function of momentum transfer, $q$ (where $q = 4 \pi \sin \theta / \lambda$, $\theta$ is the scattering angle, and $\lambda$ is the x-ray wavelength), provides the liquid structure factor, which in conjunction with the mean density can be used to determine the radial distribution function \cite{Kaplow1965,Eggert2002}.

In this paper, we present a new approach to quantitative structure factor and density determination in liquid diffraction data obtained using a pink x-ray source. The implementation of a Taylor series expansion of the spectra can be used to account for the artefacts introduced by the pink x-ray beam. Furthermore, the corrected spectra can be used in an optimization procedure to determine density of the compressed liquid state. \change{We will only focus on the case of monatomic liquid in this manuscript. The extension to the case of polyatomic liquids, although tedious, follows the same general steps.} The methods section briefly describes the scattering of x-rays by \change{monatomic} liquids. Diffraction signal modulation in the presence of a pink x-ray beam is also described. A method based on the Taylor series expansion of the coherent diffraction intensity to correct for this effect is presented. Finally, we describe the optimization scheme to derive liquid densities from data collected in a pink x-ray beam. We presents the results of the outlined procedure on two datasets: a simulated pink beam diffraction spectra derived using radial distribution function from a quantum molecular dynamics simulation and an experimental spectra recorded at the dynamic compression sector for liquid tin. The densities derived for the experimental data is compared to other experimental studies and the Sesame $2161$ EOS table. We conclude the paper by providing a brief summary and some practical considerations while using the proposed scheme.

\section{Methods\label{sec:methods}}
This section mathematically describes the main ideas of this work. The section proceeds in the following steps: a brief overview of the coherent diffraction intensity for a monatomic liquids using the Debye scattering equation is presented. Next, following the work presented in \cite{Bratos2014}, the modulation of this diffracted intensity in the presence of a pink x-ray beam is outlined. The section describes the pink beam correction using a Taylor series to approximate the change in diffraction intensity as a function of momentum transfer. Finally, the density optimization algorithm using the pink beam correction is presented.

\subsection{Monatomic Liquids}
X-ray scattering from a disordered group of $N$ atoms is given by the Debye scattering equation as \cite{Debye1915}
\begin{equation}
    I^{c}(q) = Nf(q)^2 + \sum_i^{N} \sum_{j\neq i}^{N} f_i(q) f_j(q)\frac{\sin q r_{ij}}{qr_{ij}}.
    \label{eq:Debye1}
\end{equation}
Here, $r_{ij}$ is the distance between atoms $(i,j)$, $q$ is the momentum transfer and $f$ is the atomic form factor. For monatomic liquids, the structure factor, $S(q)$ is related to the experimentally observable coherent diffraction intensity, $I^{c}$ (ignoring central scattering) by
\begin{equation}
    S(q) = \frac{I^{c}(q)}{N f^2(q)}.
    \label{eq:SQ_1}
\end{equation}
The structure factor is related to the radial distribution function, $g(r)$ by the following equation.
\begin{align}
    S(q) &= 1 + \frac{4\pi\rho}{q} \int_{0}^{\infty} \left[ g(r) -1\right]r\sin (qr) \mathrm{d}r. \nonumber \\
    &= 1 + \frac{1}{q} \int_{0}^{\infty} F(r) \sin (qr) \mathrm{d}r. 
    \label{eq:SQ2}
\end{align}
Here, $F(r) = 4\pi \rho(g(r) - 1)r$ and $\rho$ is the mean density of the liquid. Readers are referred to \cite{Warren1990} for a detailed overview of scattering by liquids.

\change{
\subsection{Structure factor from Experimental Scattering Intensities}
The experimental diffraction signal recorded from liquids can be converted to the structure factor, $S(q)$ by computing a normalization factor. The normalization factor, $\alpha$ is computed as the value which minimizes the quantity
\begin{align*}
   \sum_{q} q^2 \left [\alpha I_{expt}(q) - \mu - I_{inc}(q) - f^{2}(q)\right ].
\end{align*}
Here, $I_{expt}$ denotes the experimentally recorded diffraction intensity, $\mu$ refers to multiple scattering, I$_{inc}$ denotes the theoretically calculated incoherent intensity and $f^{2}$ refers the the squared atomic scattering factor of the liquid. $\mu$ is independent of $q$ and is another free variable along with $\alpha$ in the optimization problem. The theoretical incoherent scattering and atomic form factors for different atoms have been tabulated as a function of the scattering parameter, $s = q/4\pi$ \cite{Smith1975,IUCr}. The structure factor is obtained from the experimental diffraction intensity using the following formula
\begin{align}
    S(q) &= \frac{\alpha I_{expt}(q) - I_{inc}(q)}{f^{2}(q)}.
\end{align}
All symbols have been previously defined. The readers are referred to the classic papers of Ashcroft and Langreth for more details \cite{Ashcroft1967a,Ashcroft1967b,Ashcroft1967c}. The structure factor obtained after this is then used in an iterative loop to reduce the oscillations at small atomic distances, $r$, below the first interatomic peak of the monatomic liquid. This iteration typically converges in a few steps. The readers are referred to \cite{Eggert2002} for further details about this iterative procedure.}

\subsection{Signal Modulation by Pink X-Ray Beam\label{sec:sigmod}}
The coherent scattering signal from a liquid is modified in the presence of a pink x-ray beam. This is given by a weighted sum of the scattering by the liquid for the different energies, $E^\prime$ in the pink beam \cite{Warren1990,Bratos2014}. The weights, $w$ \change{as well as the limits of the integration, $E_{min}$ and $E_{max}$} are given by the energy spectrum produced by the undulator. Mathematically,
\begin{equation}
    I^{c}_{pink}(\theta) = \change{\int_{E_{min}}^{E_{max}}}w(E^\prime,\theta)I^{c}(E^\prime,\theta)\mathrm{d}E^\prime.
    \label{eq:pinkbeam}
\end{equation}
Here, $I^{c}$ denotes the coherent diffraction intensity. Instead of using the variable $ \theta$, it is useful to transform eq.~\ref{eq:pinkbeam} in terms of the momentum transfer, $q$. Let $E^{M}$ be some energy in the pink spectrum with non-zero photon flux. The pink coherent scattering intensity, $I^{c}_{pink}$ as a function of the scattering angle, $\theta$ can be transformed to an equivalent scattering intensity as a function of the momentum transfer variable, $q^{M}$ where $q^{M} = 4\pi E^{M}\sin\theta/hc$, where $h$ and $c$ are Planck's constant and speed of light respectively. Equation~\ref{eq:pinkbeam} transforms to
\begin{equation}
    \change{I^{c}_{pink}(q^{M}) = \int_{E_{min}}^{E_{max}}w(E^\prime, q^{M}/E^{M})I^{c}(E^\prime, q^{M}/E^{M})\mathrm{d}E^\prime.}
    \label{eq:EM}
\end{equation}
\change{Since the weights, $w$ are only dependent on the energy, its dependence on $\theta$ through the variable $Q^M/E^M$ will be dropped for all subsequent equations}. An obvious choice for $E^{M}$ would be the photon energy with the highest flux, but there is no \textit{apriori} reason for this. As we shall see later, we treat this energy as another variable to be determined during density optimization.

\subsection{Pink Beam Correction}
As discussed in section~\ref{sec:sigmod}, the liquid diffraction signal recorded in a pink x-ray beam is a linear combination of diffraction signal resulting from each energy in the x-ray. Therefore, the usual analysis methods used for monochromatic x-ray beams can't be employed directly. However, if the pink beam has a narrow energy bandwidth, as is the case at the Dynamic Compression Sector, the pink beam diffraction spectra scan be corrected to an equivalent quasi-monochromatic diffraction \change{signal}. Once this correction is performed, the known analysis methods for monochromatic liquid diffraction spectra are valid. This section outlines the procedure for performing this correction. The change in the scattering signal as a function of $q$ can be approximated by a Taylor series as
\begin{equation}
    I^{\change{c}}(q+\delta q) = I^{\change{c}}(q) + \frac{\partial I^{\change{c}}(q)}{\partial q} \delta q + \frac{\partial^2 I^{\change{c}}(q)}{\partial q^2} (\delta q)^2 + \cdots.
\end{equation}
In the limit \change{$\delta q/q \ll 1$}, this expression is well approximated by the first order term. \change{For the U17 undulator spectrum at DCS, the flux decreases by a factor of $1/e$ over the energy range $\Delta E \sim 0.88$ keV. This corresponds to a $\delta q/q \sim 0.037$, significantly smaller compared to $1$. The pink beam sources at other synchrotron sources, such as the European Synchrotron Radiation Facility (ESRF) are sharper compared to DCS ($\delta q/q \sim 0.024$) \cite{Wulff2002}}. We will make this assumption for the rest of this manuscript. Substituting the expression for $I^{c}(q)$ from equation~\ref{eq:SQ_1}, the derivative of the coherent scattering intensity is given by
\begin{align}
    \frac{\partial I^{\change{c}}}{\partial q} &= \frac{\partial}{\partial q}\left\{ N f^2(q) S(q)\right\} \nonumber \\ 
    &= N\left\{ 2f(q)\frac{\partial f(q)}{\partial q}S(q) +
    f^{2}(q)\frac{\partial S}{\partial q}\right\}. 
    \label{eq:partial_1}
\end{align}

Using the expression for the liquid structure factor from equation~\ref{eq:SQ2} results in the following expression for the derivative of the structure factor with respect to $q$
\begin{align}
    \frac{\partial S}{\partial q} =& \frac{\partial}{\partial q}\left \{1 + \frac{1}{q} \int_{0}^{\infty} F(r) \sin (qr) \mathrm{d}r \right\}. \nonumber \\ 
     =& \int_{0}^{\infty} F(r)\left( \frac{qr\cos qr - \sin (qr)}{q^2}\right) \mathrm{d}r. 
     \label{eq:derivativeSq}
\end{align}
Using the expression for the derivative of the scattering factor in equation~\ref{eq:derivativeSq} and replacing it in equation~\ref{eq:partial_1} results in the following expression for the derivative of the coherent scattering intensity
\begin{align}
    \frac{\partial I^{\change{c}}}{\partial q} = N\left\{ 2f(q)\frac{\partial f(q)}{\partial q}S(q) +  f^{2}(q)\int_{0}^{\infty} F(r)\left( \frac{qr\cos qr - \sin (qr)}{q^2}\right) \mathrm{d}r\right\}.
    \label{eq:der_Ic}
\end{align}
The form factors, $f(q)$ in the previous equations are tabulated for each atoms as a function of the parameter $s = q/4\pi$. The form factors are expressed as a weighted sum of gaussians of different widths and a constants term. Since these functions are smooth, the derivatives of the form factors are trivial to compute as well. The expression is presented in the following equations. The values of $A_i, B_i$ for different atoms have been tabulated and can be found in the international tables of crystallography \cite{IUCr}. 
\begin{align}
    f(q) &= \sum_{i=1}^{4}A_i \mathrm{e}^{-B_i s^2} + C. \nonumber \\ 
    \frac{\partial f(q)}{\partial q} &= \frac{1}{4\pi} \sum_{i=1}^{4}-2 A_i B_i s \mathrm{e}^{-B_i s^2}.
\end{align}
The derivative of the coherent scattering intensity with respect to the momentum transfer can be converted to the derivative with respect to the photon energy, $E$ by using the chain rule
\begin{equation}
    \frac{\partial I^{\change{c}}}{\partial E} = \frac{\partial I^{\change{c}}}{\partial q} \frac{\partial q}{\partial E} = \frac{q}{E} \frac{\partial I^{\change{c}}}{\partial q}.
    \label{eq:chain}
\end{equation}
Substituting the above expression in equation~\ref{eq:der_Ic} results in the following equation for the derivative of the scattering intensity with the photon energy
\begin{equation}
    \frac{\partial I^{\change{c}}}{\partial E} = \frac{N q}{E} \left\{ 2f(q)\frac{\partial f(q)}{\partial q}S(q) \right. \label{eq:dIdE} \nonumber \\ 
    + \left. f^{2}(q)\int_{0}^{\infty} F(r)\left( \frac{qr\cos qr - \sin (qr)}{q^2}\right) \mathrm{d}r\right\}.
\end{equation}
The expression for the derivative presented in the previous equation can now be plugged into equation \ref{eq:pinkbeam} to compute the modulation of the scattered x-rays as a result of the pink x-ray beam. This is given by
\change{
\begin{align}
    I^{c}_{pink}(q^{M}) &= \int_{E_{min}}^{E_{max}}w(E^\prime)I^{c}(E^\prime, q^{M}/E^{M})\mathrm{d}E^\prime  \nonumber \\ 
    &= \int_{E_{min}}^{E_{max}} w(E^\prime) \Bigg\{ I^{c}(E^{M}, q^M/E^{M}) + \frac{\partial I^{c}}{\partial E^\prime}\bigg\vert_{E^{M}}(E^\prime -E^M)\Bigg\}\mathrm{d}E^\prime \nonumber \\ 
\end{align}
Notice that $I^{c}(E^{M}, q^M/E^{M})$ and $\frac{\partial I^{c}}{\partial E^\prime}\vert_{E^{M}}$ are independent of the integration variable. This allows us to write the above equation as
\begin{align}
    I^{c}_{pink}(q^{M}) = I^{c}(E^{M}, q^M/E^{M}) + \frac{\partial I^{c}}{\partial E}\bigg\vert_{E^{M}} \int_{E_{min}}^{E_{max}} w(E^\prime) (E^\prime - E^M)\mathrm{d}E^\prime.
    \label{eq:summation}
\end{align}
}
The integral can be computed with the knowledge of the pink beam spectra. Note that for symmetric undulator spectra, this integral goes to zero. Therefore, in the first order approximation, the correction term goes to zero for a symmetric profile. However, this is not true if higher terms are included in the Taylor series. The effective monochromatic spectra can be computed by subtracting the correction term from the pink beam spectra. The effective monochromatic spectrum is given by
\begin{equation}
    I^{\change{c}}(E^M) = I^{c}_{pink}(q^{M}) - \frac{\partial I^{\change{c}}}{\partial E}\bigg\vert_{E^{M}} \change{\int_{E_{min}}^{E_{max}} w(E^\prime)} (E^\prime - E^M)\mathrm{d}E^\prime.
    \label{eq:monotopink}
\end{equation}
The above equation is only valid for a \change{monatomic} liquid. A similar correction for polyatomic liquid \change{can also be calculated using the formalism described above but is beyond the scope of this work.}

\change{
\subsection{Termination Function}
To partially eliminate effects of limited $q$-range in the measured signal, the termination function described in \cite{Kuwayama2020} was used. This method extends the structure factors beyond the recorded limit, $q_{\rm max}$ using the following equation
\begin{align}
    S_{\textrm{extend}}(q) = \begin{cases} S(q),\quad & q \leq q_{\textrm{max}} \\
    1 - \frac{1}{q}\int_{0}^{r_{\textrm{cutoff}}}(4\pi\rho_{N}+\frac{2}{\pi}\int_{0}^{q_{\textrm{max}}}q(S(q) - 1)\sin (qr)~\mathrm{d}q)\mathrm{d}r, \quad & q > q_{\textrm{max}}.
    \end{cases}
\end{align}
Here, $\rho_{N}$ refers to the number density. The other symbols have previously been defined in the text.
}
\subsection{Density Optimization}
\begin{figure}
    \centering
    \includegraphics[width=\columnwidth]{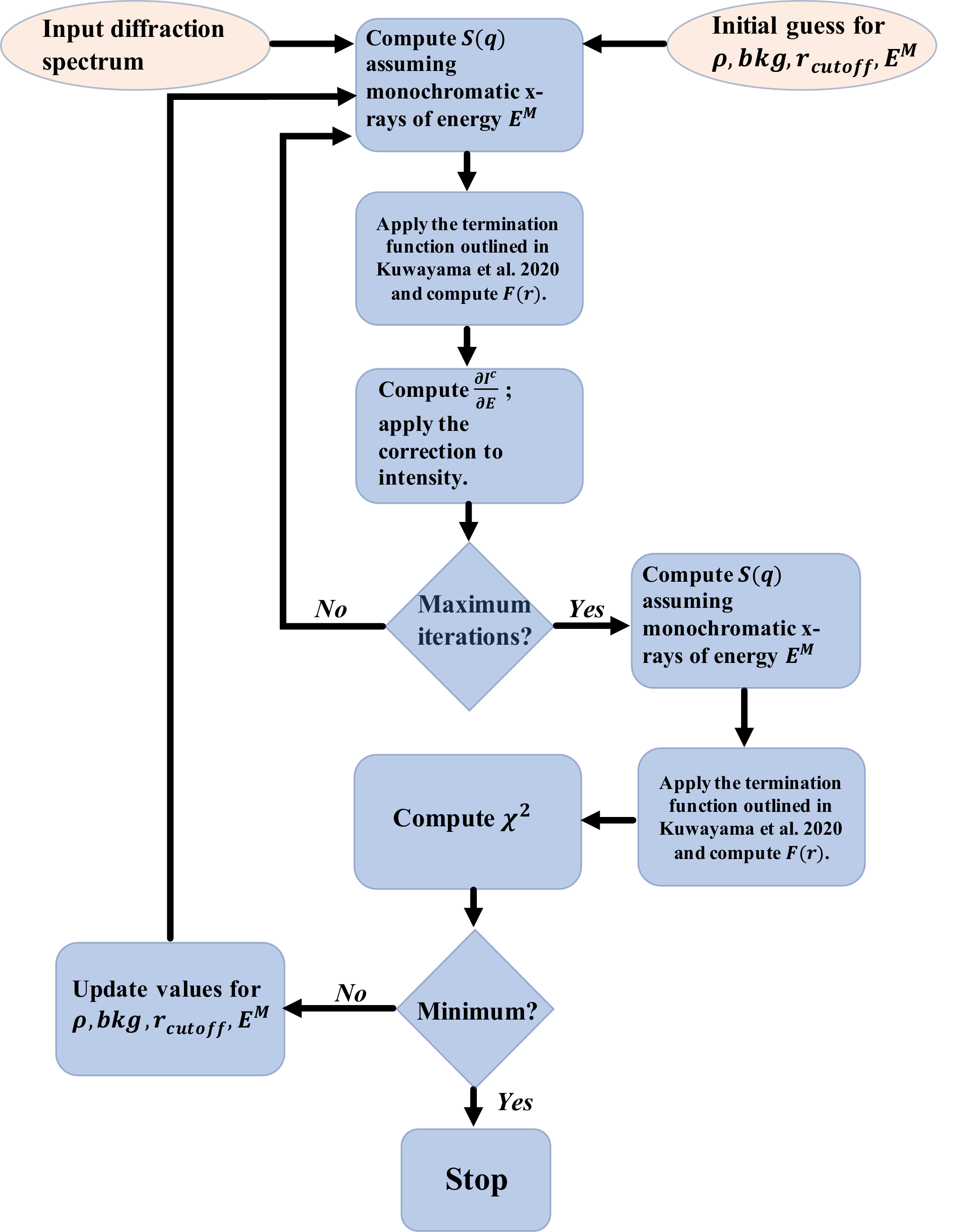}
    \caption{Density optimization algorithm. Symbols have been defined in the text.}
    \label{fig:flowchart}
\end{figure}

The above formalism assumes the knowledge of $F(r)$ to compute the derivative of the coherent scattering intensity. However, this information is not known apriori and needs to be extracted from the measured diffraction intensity. Therefore, a bootstrap method was used. This is outlined in the algorithm below
\begin{enumerate}
    \item Compute $g(r)$ (\change{inverse Fourier transform of} eq.~\ref{eq:SQ2}) without any correction and assuming monochromatic spectrum using measured signal.
    \item Compute correction factor using eq.~\ref{eq:dIdE} and $g(r)$ from 1.
    \item Use correction factor from 2, correct for pink beam using eq.~\ref{eq:monotopink} and recompute $g(r)$ .
    \item Use the corrected $g(r)$ to update the current values for input parameters.
    \item Repeat steps $2-4$ until converged.
\end{enumerate}
It should be noted that due to limited $q$-range of the measured signal, iterative application the correction \change{can lead to growing fluctuation in $S(q)$. This is related to the rapidly oscillating nature of the derivative of the $\textrm{sinc}$ function (second term on the right in eq.~\ref{eq:der_Ic}) }. Practically, only one iteration leads to acceptable results and avoids numerical instability. The corrected structure factor is then fed into a similar optimization procedure to the one outlined in \cite{Eggert2002} to extract the density. \change{The optimization problem seeks to minimize the following function
\begin{align}
    \chi^{2} = \int_{r_0}^{r_{cutoff}}(F(r) + 4\pi\rho_{n})^{2}\mathrm{d}r
\end{align}
Here, $r_{cutoff}$ denotes the interatomic distance below which the radial distribution function should vanish, $\rho_{n}$ and $\change{bkg}$ define the number density in atoms/$\AA^{-3}$ and constant background signal respectively and $E^{M}$, defined in section \ref{sec:sigmod} denotes the energy in the Taylor series approximation. Other symbols have been previously defined. The optimization is performed over $4$ variables, namely density, background signal, $r_{\rm cutoff}$ and $E^{M}$}. The optimization parameters are updated using the BOBYQA optimization algorithm \cite{bobyqa,pybobyqa}. Fig.~\ref{fig:flowchart} outlines the density optimization algorithm. \change{Note that the algorithm accepts bounds constraints on variables. In our experience, a constraints of $\pm 0.2-0.3 \AA$ for $r_{cutoff}$ around an initial guess derived from QMD simulations and $\pm 1$ keV around an initial guess of peak flux energy for E$^M$ are a good choice.} 



\section{Results\label{sec:results}}
This section presents results of the outlined procedure for two different cases: a simulated pink beam diffraction \change{signal} derived from quantum molecular dynamics simulation and experimental diffraction \change{signal} recorded at the Dynamic Compression Sector. The simulated results have a known ground truth and provides a good measure for the efficacy of the proposed method. The density derived from the experimental \change{diffraction signal} is compared to densities derived from other shock compression and gas gun studies as well as the seasame 2161 pressure-density table.

\subsection{Quantum Molecular Dynamics simulations - Tin\label{sec:QMD}}
 Quantum molecular dynamics (QMD) simulations based on density functional theory were performed for liquid Sn using a cubic cell with 128 atoms. We use the Baldereschi $k$ point \cite{Baldereschi1973} of $(\frac{1}{4} ; \frac{1}{4} ; \frac{1}{4})2\pi/a$, to sample the Brillouin zone. Here, $a$ is the side length of the simulation cell. We choose the Perdew-Burke-Ernzerhof for solids (PBEsol) exchange-correlation functional \cite{Blochl1994}, $400$ eV cutoff for the plane wave basis, and a projected augmented wave (PAW) pseudopotential that has a core of 3.0 Bohr and treats $5s^25p^2$ as valence electrons as provided in the Vienna Ab-Initio Simulation Package (VASP) \cite{VASP}. A Nos\'{e} thermostat was used to generate MD trajectories in a canonical (NVT, i.e., constant number of atoms, constant volume, and constant temperature) ensembles. The MD trajectory consisted of $12 000$ steps with a time step of approximately 1.5 fs. The radial distribution function is calculated by analyzing interatomic distances along the MD trajectory after the system reaches equilibrium, from which the structure factor is calculated following the procedure outlined in \cite{Zhang2020}. We note that the simulation being reported here corresponds to a temperature of 5000 K. We have performed additional calculations at $\pm$ 1000 K, using different cell sizes (up to 256 atoms), finite $k$ mesh, and other exchange-correlation functionals, and found similar results for the calculated pressure and $g(r)$ profile.
 
\begin{figure}
    \centering
    \includegraphics[width=\textwidth]{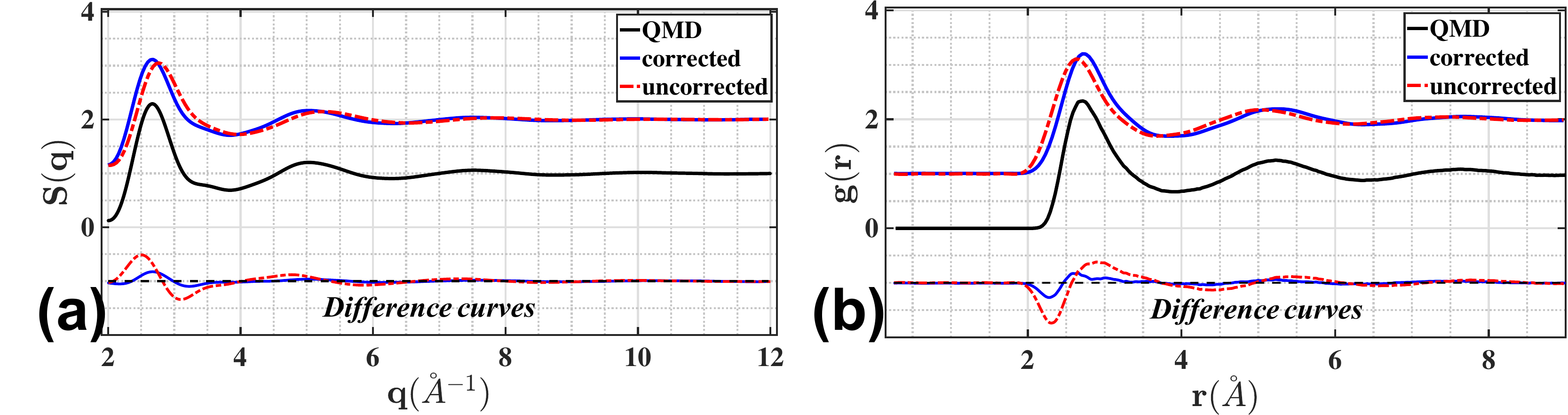}
    \caption{(a) Structure factor and (b) radial distribution function for tin derived from QMD simulations at 11.0 g/cc and 5000 K. The finite size of the QMD simulation cell results in the structure factor below $\sim 2\AA^{-1}$ being unreliable. The difference curve of the corrected and uncorrected curves with the QMD results is shown as well.}
    \label{fig:5000K}
\end{figure}

The radial distribution function was extracted from QMD simulations and equations \ref{eq:SQ_1}, \ref{eq:SQ2} and \ref{eq:pinkbeam} were used to generate the ``experimental" \change{diffraction signal} \change{for pink x-ray beam}. The energy-flux distribution recorded for the $U17$ undulator at DCS as shown in Fig.~\ref{fig:xrays} was used for this calculation. The procedure outlined in this work was then used to extract the structure factor, radial distribution function and mean density. \change{This is represented by the ``corrected" curved in Fig.~\ref{fig:5000K}(a)-(b). The ``uncorrected" curves does not account for the non-monochromatic source and assumes that the coherent diffraction signal was recorded at a monochromatic source with x-ray energy corresponding to the energy with the peak flux in Fig.~\ref{fig:xrays} (red curve $\sim 23.53$ keV)}. This data set lets us benchmark the algorithm for the ideal case where the mean density is already known. 
The theoretical density (in $\textrm{g cm}^{-3}$) along with the densities obtained for the uncorrected and corrected cases are listed in table~\ref{tab:A}. The $r_{cutoff}$ and $E^{M}$ converged to the physically reasonable values of $2 \AA$ and $23.48$ keV respectively. Unlike the Levenberg-Marquardt least-squares algorithm \cite{Watson1978}, the BOBYQA algorithm used for this work does not give uncertainties in the estimated parameters using the covariance matrix. However, this can be estimated by starting the optimization at different initial parameter values. The algorithm converges to different values for the different starting values. The mean and the standard deviation of the converged values of density can be used as the mean value and the uncertainty in the density. Similar values can also be derived for the other parameters such as the background scaling, cutoff radius etc. The starting density was uniformly sampled in the interval $10.5-11.5~\textrm{g cm}^{-3}$. The uncertainties reported in table~\ref{tab:A} are the standard deviation for $100$ different starting values. It is interesting to note that the uncertainties for the corrected density is not large enough to account for the deviation of the estimated density with the theoretical value. This indicates that that although the approximation clearly improves the estimate, the Taylor series approximation is not exact and introduces has a small ($<1\%$) error in the final estimate. This is likely a result of the first order approximation made in this work. However, including higher order terms is infeasible due to the limited $q-$range in experimental data. This introduces unwanted fluctuations in the corrected structure factors and leads to erroneous estimates of density. The QMD results for the structure factor and radial distribution functions, along with the uncorrected and corrected spectra for these quantities are presented in Fig.~\ref{fig:5000K} respectively. 

\begin{table}
\centering
 \begin{tabular}{|| c | c | c ||} 
 \hline
Theoretical density & Uncorrected density & Corrected density  \\ 
 \hline
 $11.0$ & $12.250(2)$ & $10.92(1)$ \\
  & (6.90 \%) & (0.73 \%) \\
 \hline
\end{tabular}

\caption{\label{tab:A} Theoretical mean density and densities obtained by the outlined optimization procedure before and after correcting for the pink beam effect. Percentage errors are shown in parentheses. All densities in units of $\textrm{g cm}^{-3}$.}
\end{table}
\subsection{Experiment - Tin\label{sec:ExptTin}}
Laser shock compression experiments were carried out at the Dynamic Compression Sector of the Advanced Photon Source synchrotron, Argonne National Laboratory \cite{DCS}. Shock targets consisted of 50 $\mu$m of polyimide ablator, glued to 29.5 $\mu$m Sn foils (99.75\% Goodfellow). A 500 $\mu$m thick single crystal LiF window was glued to the rear surface of the target package and velocimetry measurements were collected at the Sn/LiF interface using a point VISAR system. Pressure was determined by impedance matching the measured Sn/LiF particle velocity and comparing to the Sn EOS (Sesame 2161) \cite{Seasame}.
\begin{figure}
    \centering
    \includegraphics[width=\textwidth]{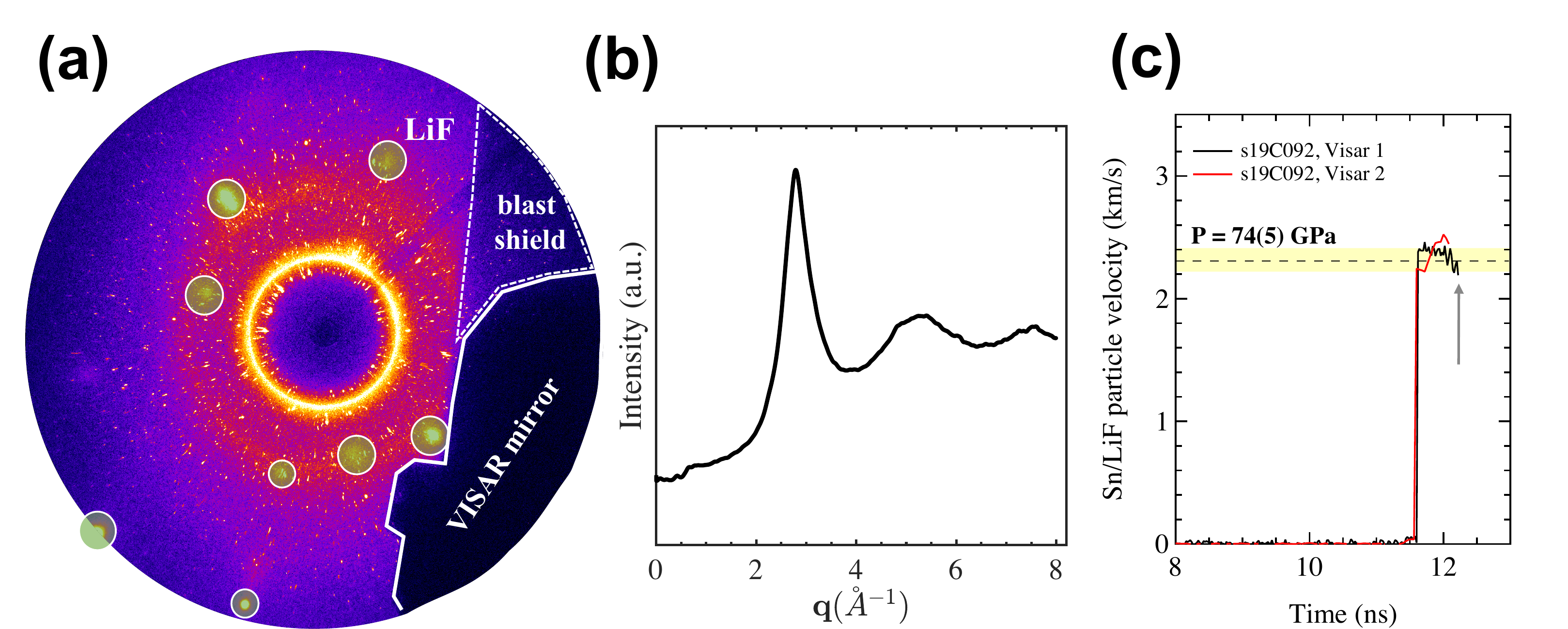}
    \caption{(a) Raw diffraction profile collected at DCS. Single crystal spots from the LiF window material, which are masked from the integrated profiles, are highlighted by green circles. All remaining spots are ambient $\beta$-Sn peaks (ambient material ahead of the shock wave) that are also removed from the final integrated profile, leaving only the diffuse scattering from liquid tin. The x-ray shadows from the VISAR mirror and blast shield are also highlighted. (b) partial azimuthally integrated intensity as a function of momentum transfer from (a) ignoring the areas with shadows from the blast shield and the VISAR mirror. The beam energy used for converting from $2\theta$ to $\mathbf{q}$ was $23.53$ keV. (c) point VISAR data. The dashed line shows the average Sn/LiF particle velocity at shock breakout, with uncertainty bounded by the yellow shaded region; the shock pressure of 74(5) GPa is determined using impedance matching of the Sn sample with the LiF window. The arrow indicates the time at which there is a significant loss of reflectivity.}
    \label{fig:Ipink}
\end{figure}
X-ray images were collected on a Rayonix SX165 area detector (2048 x 2048 pixels) and timed with respect to the laser pulse such that diffraction was collected just before the shock wave reaches the Sn/LiF interface. \change{Consequently, a small portion of ambient diffraction from uncompressed Sn ahead of the shock wave is masked from diffraction profiles so that only liquid scattering signal is considered for analysis. Single crystal Laue diffraction spots from the single crystal LiF window are also masked. Finally, the detector response is removed from the shocked images before azimuthally integrating using Dioptas \cite{dioptas}. This approach was shown to work well for data collected at the LCLS \cite{Briggs2019} and at DCS \cite{Briggs2019b}. However, the contribution of background signal from the plastic ablator will be more significant for lower Z materials and would require additional background subtraction, similar to the removal of empty cell background as described in \cite{Eggert2002}. In this work the ratio of the squares of atomic numbers, representing the scattering cross-section for Sn and Kapton (CH) is $\sim 100$, and the contribution of the $50 \mu$m plastic ablator is negligible (we observe no amorphous or crystalline signal from the plastic ablator) at $\sim 23.5$ keV.}

Consequently, a small portion of ambient diffraction from uncompressed Sn ahead of the shock wave is masked from diffraction profiles so that only liquid scattering signal is considered for analysis. Single crystal Laue diffraction spots from the single crystal LiF window are also masked. \change{The readers are referred to \cite{Briggs2019} for details about the background removal procedure}. See Figure \ref{fig:Ipink}(a). The detector response was removed from the shocked images, before azimuthally integrating using Dioptas \cite{dioptas}. 

The sample detector distance, rotation, and tilt, were calibrated using the diffraction lines of polycrystalline Si and cross checked with CeO2 (NIST). The azimuthally integrated intensity after removing the diffraction from crystalline Sn and LiF window as well as the shadowed regions from the blast shield and the VISAR mirror is shown in Fig.~\ref{fig:Ipink}(b). A pressure of $74(5)$ GPa was estimated from VISAR as shown in Fig.~\ref{fig:Ipink}(c). Similar experiments were performed on the Matter in Extreme Conditions (MEC) instrument using the monochromatic x-ray Free Electron Laser at the Linac Coherent Light Source (LCLS-II) to achieve similar densities. The results have been previously reported in \cite{Briggs2019}. The study reported a mean density of $11.2(1)~\textrm{g cm}^{-3}$. This corresponds to an estimated pressure of $79(8)$ GPa on the Sesame 2161 Hugoniot. \change{Earlier work using multi-anvil apparatus studied liquid Tin only up to $20$ GPa \cite{Narushima2007} at temperatures just above the melting curve \cite{Briggs2017b}.}

\begin{figure}
    \centering
    \includegraphics[width=\textwidth]{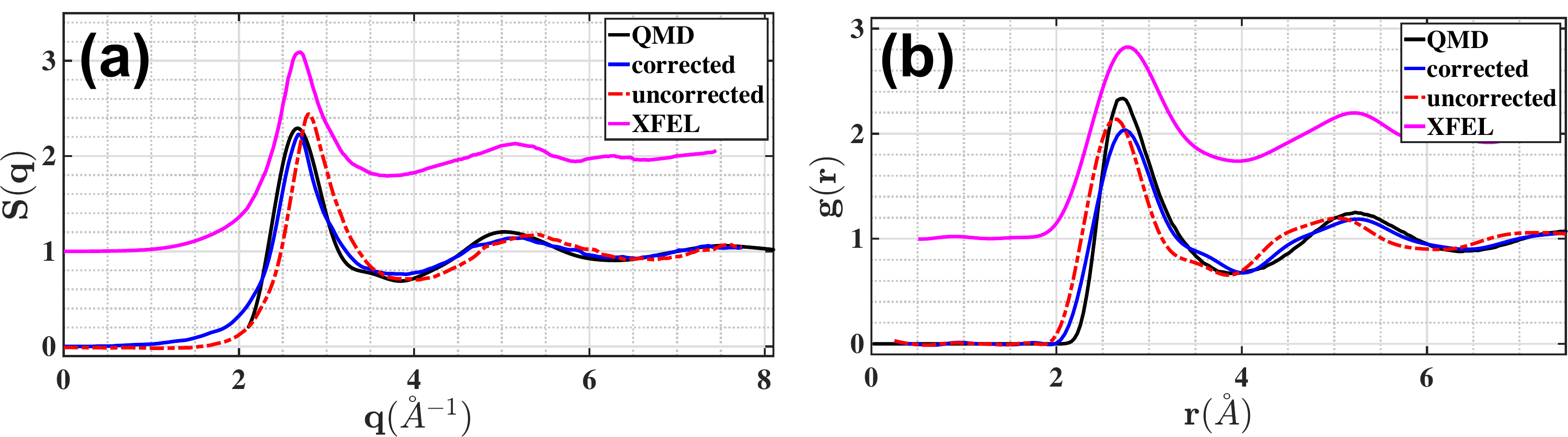}
    \caption{(a) Structure factor and (b) radial distribution function from shock melted tin for measurements at LCLS-II, DCS (corrected and uncorrected) as well as the QMD simulations. While a direct comparison is not possible (see text), the proposed algorithm leads to reasonable results for liquid tin of comparable pressures and densities.}
    \label{fig:DCS_Sn}
\end{figure}

Density optimization without accounting for the pink beam resulted in a mean density of $ 13.67(7)~ \textrm{g cm}^{-3}$. Accounting for the pink beam effect resulted in a density of $11.0(2)~ \textrm{g cm}^{-3}$. The QMD simulations from the previous example in Sec.~\ref{sec:QMD} had a pressure of $80$ GPa, comparable to the value estimated from VISAR analysis. The structure factors and the radial distribution functions obtained using the data recorded at DCS and MEC along with the QMD simulations are presented in Fig.~\ref{fig:DCS_Sn}. Since the temperatures are not measured during experiments, the precise thermodynamic state of the melted tin is not known. This makes the direct comparison between the two experimental measurements and the QMD simulations difficult. The purpose of presenting these datasets is to not to do a quantitative comparison but to demonstrate that the algorithm outlined in this work leads to reasonable estimates of the radial distribution functions for liquid tin of comparable densities. The coordination number of the corrected spectra, related to the area under the first $g(r)$ peak, shows agreement with the QMD simulations and the MEC data once the correction is applied. \change{The coordination number was determined using the following equation \cite{Morard2014}
\begin{equation}
    CN = \int_{r_0}^{r_{min}}4\pi nr^{2}g(r)\mathrm{d}r.
    \label{eq:CN}
\end{equation}
Here, $n$ is the number density and $r_{0}$ and $r_{min}$ are the integration limits corresponding to the left edge of the first peak and the first minima in $g(r)$ respectively. There are competing methods prescribed to evaluate $r_{min}$, where $r_{min}$ corresponds to the first minima of the function $4\pi r^{2}g(r)$ \cite{waseda1980}. This convention was used by the the authors in \cite{Briggs2019} to compute the coordination number. Table~\ref{tab:B} lists the coordination number obtained by using the definition of $r_{min}$ in \cite{Morard2014} (method I) and \cite{waseda1980} (method II). The values indicate that using the same cutoff value for $r_{min}$ leads to consistent results. }

\begin{figure}
    \centering
    \includegraphics[width=\columnwidth]{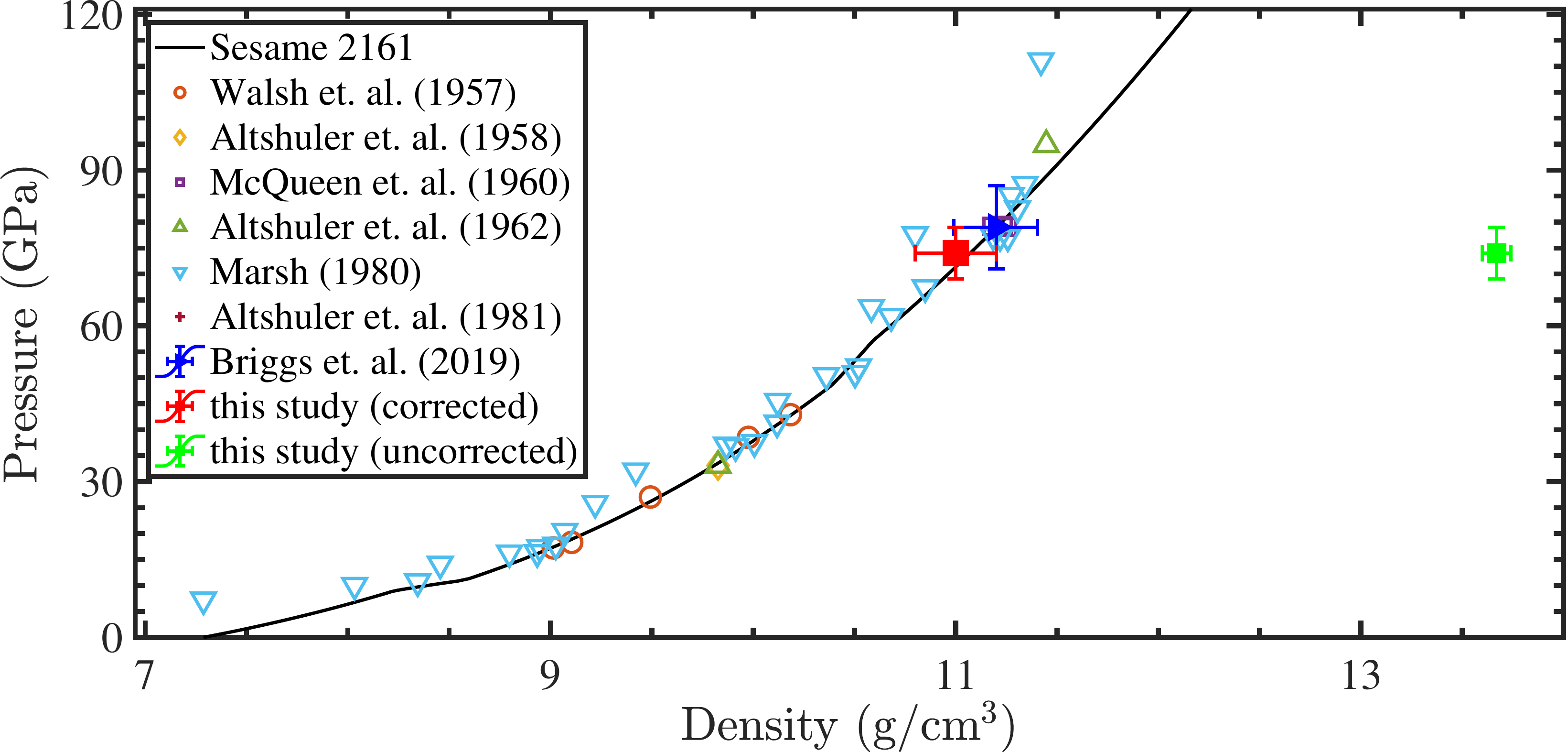}
    \caption{Tin densities on the Hugoniot derived from gas gun experiments \change{\cite{Walsh1957,Altshuler1958,McQueen1960,Altshuler1962,Marsh1980,Altshuler1981}} with density derived from current study plotted with red glyph. The SESAME 2161 pressure-density table is shown with the solid black line \cite{Seasame}. The density without accounting for the pink beam is shown in the green glyph. \change{It should be noted that the reported liquid density using dynamic compression experiments in \cite{Briggs2019} was estimated by pinning the VISAR pressure on the Sesame 2161 hugoniot. Quantitative estimate was not possible using the collected diffraction signal.}}
    \label{fig:GasGun}
\end{figure}

Due to the higher density estimate, the uncorrected radial distribution function overestimates the coordination number significantly. \change{The authors in \cite{Zhang2020} use method II to compute the CN and argue that since the body centered cubic phase of tin has $8$ first nearest neighbors and $6$ second nearest neighbors, a coordination number close to $14$ is reasonable for the liquid phase, and indicates that these two shells merge into one.} These results are summarized in Table~\ref{tab:B}. \change{The uncertainty in the mean density for both the uncorrected and corrected spectra is estimated by initializing the optimization at $100$ uniformly sampled density values in the interval $10.5-11.1~\textrm{g cm}^{-3}$. }
\change{
\begin{table}
\centering
 \begin{tabular}[0.8\textwidth]{||c | c | c | c | c||} 
 \hline
Data Source & Pressure (GPa) & Density ($\textrm{g cm}^{-3}$) & CN (method I) & CN (method II)  \\ 
 \hline
 DCS (uncorrected) & $74(5)$ & $13.67(7)$ & $15.5(1)$ & $13.2(1)$\\
  \hline
 DCS (corrected) & $74(5)$ & $11.0(2)$ & $13.5(2)$ & $11.4(2)$\\
  \hline
 XFEL @ LCLS-II & $79(8)$ & $11.2(1)$ & $13.7(1)$ & $11.1(1)$\\
  \hline
 QMD Simulations & $79.95(8)$ & $11.0$ & $13.6$ & $12.0$\\
 \hline
\end{tabular}
\caption{\label{tab:B} Densities and coordination numbers for liquid tin from different datasets. Two different methods used for calculating the CN, as discussed in section~\ref{sec:ExptTin}, have been presented as method I and II.}
\end{table}
}
To put our density estimates in a broader context, the estimates densities were compared to other experimental data points as well as the tabulated sesame 2161 Hugoniot. This has been presented in Fig.~\ref{fig:GasGun}. The density estimate, after applying the pink beam correction, agree very well with previously measured data as well as the sesame 2161 tables for Tin. The density estimate without accounting for the pink beam artifacts, shown in the green glyph, is extremely poor and shows the significant impact of the pink beam on the density estimates.

\section{Discussion and Conclusions\label{sec:discussion}}
In this paper, we have outlined a new methodology to correct for the artifacts introduced in diffraction signal produced by liquids in a pink x-ray beam. The correction relies on the first order Taylor's series expansion of the coherent  diffraction intensity. The proposed method is bench-marked with a simulated tin data of known density and temperature. The method was able to correct for the pink beam effect to less than one percent error. Finally, the method was demonstrated for an experimental liquid scattering \change{signal} from tin recorded at the Dynamic Compression Sector. The mean density and radial distribution function compares favourably with both experimental results recorded with a monochromatic source at the Linac Coherent Light Source as well as quantum molecular dynamics simulations. The density estimates after applying the correction is in excellent agreement with other experimentally recorded data as well as the sesame 2161 tables for Tin. 
The above treatment is a practical one as there are no theoretical guarantees that the algorithm will converge. However, in most scenarios the methodology is able to correct for the measurement artifacts introduced by the pink x-ray beam. Furthermore, introduction of new variables in the optimization problem makes it harder to find the global minima. Care must be taken in choosing the initial starting values and the bounds specified. Quantum molecular dynamics can be used to guide the starting values and expected ranges of these parameters. Alternatively, a more robust global optimization algorithm can be utilized. This comes at an increased computational cost. Finally, a limited $q$-range in the measured data can introduce unwanted fluctuations in the corrected structure factor and radial distribution function. Therefore, the results need to be evaluated with caution to ensure that such a fluctuation is not interpreted as a true feature in the data.


\ack{
We thank Pinaki Das, Yuelin Li, Paulo Rigg, Adam Schuman, Nicholas Sinclair, Xiaoming Wang, and Jun Zhang at DCS for their assistance during laser experiments. The research was supported by the Laboratory Directed Research and Development Program at LLNL (project no. 18-ERD-012). This work of was performed under the auspices of the U.S. Department of Energy at Lawrence Livermore National Laboratory under Contract DE-AC52-07NA27344 (LLNL-JRNL-819849). This publication is based upon work performed at the Dynamic Compression Sector, which is operated by Washington State University under the U.S. Department of Energy  (DOE)/National Nuclear Security Administration award no. DE-NA0003957.  This research used resources of the Advanced Photon Source, a DOE Office of Science User Facility operated for the DOE Office of Science by Argonne National Laboratory under contract no. DE-AC02-06CH11357.}

\bibliographystyle{iucr} 
\bibliography{references.bib}  

\end{document}